\documentclass{article}%
\usepackage{amsmath, amsthm, amssymb}
\usepackage{pslatex}
\usepackage{amsfonts}
\usepackage{verbatim}
\usepackage{epsfig, graphicx}
\usepackage{amsmath}
\usepackage{amssymb}
\usepackage{graphicx}%
 \theoremstyle{plain}
\newtheorem{theorem}{Theorem}

\theoremstyle{remark}

\begin{document}
\title{Supplement to the article ``Approximating multi-dimensional Hamiltonian flows by billiards":\\
Proof of C$^{0}$ and C$^{r}$ - closeness Theorems}
\author{by A. Rapoport, V.
Rom-Kedar~~and~~D. Turaev } \maketitle

\begin{theorem}
\label{Main th0} Let the potential $V(q;\epsilon)$ in the equation
\begin{equation}
H=\frac{p^{2}}{2}+V(q;\epsilon), \label{Hamiltonian epsilon}%
\end{equation}
satisfy Conditions I-IV stated above. Let $h_{t}^{\epsilon}$ be
the Hamiltonian flow defined by (\ref{Hamiltonian epsilon}) on an
energy surface $H=H^{\ast}<\mathcal{E}$, and
$b_{t}$ be the billiard flow in $D$. Let $\rho_{0}$ and $\rho_{T}=b_{T}%
\rho_{0}$ be two inner phase points\footnote{Hereafter, $T$ always
denotes a\emph{ finite }number.}. Assume that on the time interval
$[0,T]$ the billiard trajectory of $\rho_{0}$ has a finite number
of collisions, and all of them are either regular reflections or
non-degenerate tangencies. Then
$h_{t}^{\epsilon}\rho_{\overset{\longrightarrow}{_{\epsilon\rightarrow0}}%
}b_{t}\rho$, uniformly for all $\rho$ close to $\rho_{0}$ and all
$t$ close to $T$.
\end{theorem}

\begin{theorem}
\label{Main thr} In addition to the conditions of Theorem
\ref{Main th0}, assume that the billiard trajectory of $\rho_{0}$
has no tangencies to the boundary on the time interval $[0,T]$.
Then $h_{t}^{\epsilon}~_{\overset
{\longrightarrow}{_{\epsilon\rightarrow0}}}b_{t}$ in the
$C^{r}$-topology in a small neighborhood of $\rho_{0}$, and for
all $t$ close to $T$.
\end{theorem}

\begin{proof}
\label{main} By Condition I the Hamiltonian flow is $C^{r}$-close
to the billiard flow outside an arbitrarily small boundary layer.
So we will concentrate our attention on the behavior of the
Hamiltonian flow inside such a layer.

Let the initial conditions correspond to the billiard orbit which
hits a boundary surface $\Gamma_{i}$ at a (non-corner) point
$q_{c}$. By Condition IIa, the surface $\Gamma_{i}$ is given by
the equation $Q(q;0)=Q_{i}$, hence the boundary layer near
$\Gamma_{i}$ can be defined as $N_{\delta
}=\{|Q(q;\epsilon)-Q_{i}|\leq\delta\}$, where $\delta$ tends to
zero sufficiently slowly as $\epsilon\rightarrow+0$. Take
$\epsilon$ sufficiently
small. The smooth trajectory enters $N_{\delta}$ at some time $t_{in}%
(\delta,\epsilon)$ at a point $q_{in}(\delta,\epsilon)$ which is
close to the collision point $q_{c}$ with the velocity
$p_{in}(\delta,\epsilon)$ which is close to the initial velocity
$p_{0}$. See Figure \ref{fig:hamiltonian}. The same trajectory
exits from $N_{\delta}$ at the time $t_{out}(\delta,\epsilon)$ at
a point $q_{out}(\delta,\epsilon)$ with velocity
$p_{out}(\delta,\epsilon )$. In these settings, the theorems are
equivalent ($r=0$ corresponds to Theorem \ref{Main th0}, while
$r>0$ corresponds to Theorem \ref{Main thr}) to
proving the following statements:%
\begin{equation}
\lim_{\delta\rightarrow0}\lim_{\epsilon\rightarrow+0}\bigg\|\bigg(q_{out}%
(\delta,\epsilon),t_{out}(\delta,\epsilon)\bigg)-\bigg(q_{in}(\delta
,\epsilon),t_{in}(\delta,\epsilon)\bigg)\bigg\|_{C^{r}}=0, \label{Travel is 0}%
\end{equation}
which guarantees that the trajectory does not travel along the
boundary, and
\begin{equation}
\lim_{\delta\rightarrow0}\lim_{\epsilon\rightarrow+0}\bigg\|p_{out}%
(\delta,\epsilon)-p_{in}(\delta,\epsilon)+2n(q_{in})\langle p_{in}%
(\delta,\epsilon),n(q_{in})\rangle\bigg\|_{C^{r}}=0, \label{Correct normal}%
\end{equation}
where $p_{out}=p_{in}-2\langle p_{in},n(q)\rangle n(q)$ and $n(q)$
is the unit inward normal to the level surface of $Q$ at the point
$q$.

\begin{figure}[ptb]
\centering \psfig{figure =
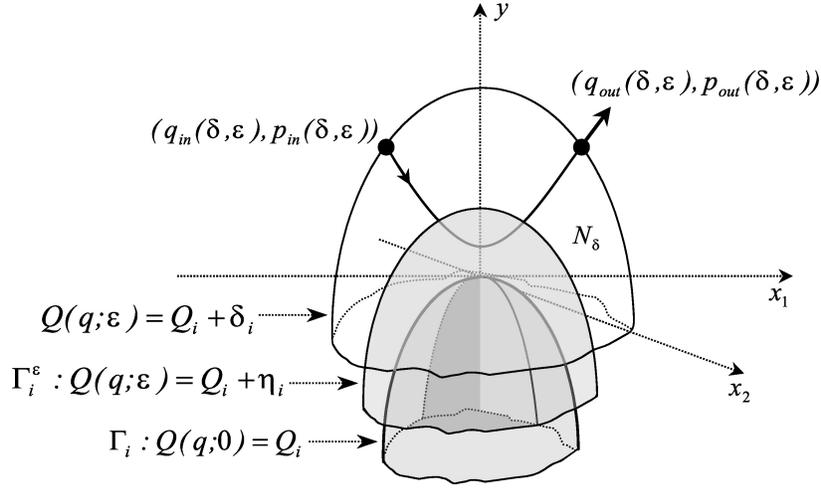,height=70mm,width=110mm}\caption{Hamiltonian flow
inside small
boundary layer.}%
\label{fig:hamiltonian}%
\end{figure}

With no loss of generality, assume that $Q(q;0)$ increases as $q$
leaves $D^{\prime}s$ boundary towards $D^{\prime}s$ interior.
Choose the coordinates $(x,y)$ so that the hyperplane $x$ is
tangent to the level surface $Q(q;\epsilon)=Q(q_{c};\epsilon)$ and
the $y$-axis is the inward normal to
this surface at $q=q_{c}$. Hence, the partial derivatives of $Q$ satisfy:%
\begin{equation}
Q_{x}|_{(q_{c};\epsilon)}=0,\;\;\;\;Q_{y}|_{(q_{c};\epsilon)}=1
\label{Partial der of Q}%
\end{equation}
By (\ref{Hamiltonian epsilon}) and Condition II, near the boundary
the equations of motion have the form:
\begin{equation}
\dot{x}=\frac{\partial H}{\partial p_{x}}=p_{x}\;\;\;\;\;\;\;\dot{p_{x}%
}=-\frac{\partial H}{\partial x}=-W^{\prime}(Q;\epsilon)Q_{x},
\label{Equations of motion x}%
\end{equation}%
\begin{equation}
\dot{y}=\frac{\partial H}{\partial p_{y}}=p_{y}\;\;\;\;\;\;\;\dot{p_{y}%
}=-\frac{\partial H}{\partial y}=-W^{\prime}(Q;\epsilon)Q_{y}.
\label{Equations of motion y}%
\end{equation}
We start with the $C^{0}$ version of (\ref{Travel is 0}) and
(\ref{Correct normal}). First, we will prove that given a
sufficiently slowly tending to zero $\xi(\epsilon)$, if the orbit
stays in the boundary layer $N_{\delta}$ for all $t\in\lbrack
t_{in},t_{in}+\xi]$, then in this time
interval%
\begin{equation}
q(t)=q_{in}(\delta,\epsilon)+O(\xi), \label{Approx for q}%
\end{equation}%
\begin{equation}
p_{x}(t)=p_{x}(t_{in}(\delta,\epsilon))+O(\xi), \label{Approx for p}%
\end{equation}%
\begin{equation}
\frac{p_{y}(t)^{2}}{2}+W(Q(q(t);\epsilon);\epsilon)=\frac{p_{y}(t_{in}%
(\delta,\epsilon))^{2}}{2}+W(\delta;\epsilon)+O(\xi).
\label{Approx for conserv}%
\end{equation}
Note that (\ref{Approx for q}) follows immediately from
(\ref{Equations of motion x})-(\ref{Equations of motion y}) and
the fact that
$p$ is uniformly bounded by the energy constraint $\frac{p^{2}}{2}%
=H-W(Q;\epsilon)\leq H=\frac{1}{2}$. In fact, $q_{in}-q_{c}$ tends
to zero as $O(\delta)$ for regular trajectories and
$O(\sqrt{\delta})$ for non-degenerate tangent trajectories, so by
assuming that $\xi(\epsilon)$ is slow enough, we extract from
(\ref{Approx for q}) that
\begin{equation}
q(t)=q_{c}+O(\xi). \label{afq}%
\end{equation}
Now, from (\ref{Partial der of Q}), (\ref{afq}), for $t\in\lbrack
t_{in}(\delta,\epsilon),t_{in}(\delta,\epsilon)+\xi]$ we have
\begin{equation}
Q_{x}(q(t);\epsilon)=O(\xi),\;\;Q_{y}(q(t);\epsilon)=1+O(\xi).
\label{Q - partial der}%
\end{equation}
Divide the interval $I=[t_{in},t_{in}+\xi]$ into two sets: $I_{<}$
where $|W^{\prime}(Q;\epsilon)|<1$ and $I_{>}$ where
$|W^{\prime}(Q;\epsilon)|\geq
1$. In $I_{<}$ we have $\dot{p_{x}}=O(\xi)$ by (\ref{Equations of motion x}%
),(\ref{Q - partial der}). In $I_{>}$, as
$|W^{\prime}(Q;\epsilon)|\geq1$ and $Q_{y}\neq0$, we have that
$\dot{p_{y}}$ is bounded away from zero, so in (\ref{Equations of
motion x}) we can divide $\dot{p_{x}}$ by $\dot{p_{y}}$:
\[
\frac{dp_{x}}{dp_{y}}=\frac{Q_{x}}{Q_{y}}.
\]
It follows that the change in $p_{x}$ on $I$ can be estimated from
above as $O(\xi^{2})$ (the contribution from $I_{<}$) plus
$O(\xi)$ times the total variation in $p_{y}$. Thus, in order to
prove (\ref{Approx for p}), it is enough to show that the the
total variation in $p_{y}$ on $I$ is uniformly bounded. Recall
that $p_{y}$ is uniformly bounded ($|p_{y}|\leq1$ from the energy
constraint) and monotone (as $W^{\prime}(Q)<0$ and $Q_{y}>0$, we
have $\dot{p}_{y}>0$, see (\ref{Equations of motion y}))
everywhere on $I$, so its total variation is uniformly bounded
indeed. Thus, (\ref{Approx for p}) is proven. The approximate
conservation law (\ref{Approx for conserv}) follows
now from (\ref{Approx for p}) and the conservation of $H=\frac{p_{y}^{2}}%
{2}+\frac{p_{x}^{2}}{2}+W(Q(q;\epsilon);\epsilon)$.

Finally, we prove that $\tau_{\delta}$, the time the trajectory
spends in the boundary layer $N_{\delta}$, tends to zero as
$\epsilon\rightarrow0$. This step completes the proof of Theorem
\ref{Main th0}: by plugging the time $\tau_{\delta}\rightarrow0$
instead of $\xi$ in the right-hand sides of (\ref{Approx for
q}),(\ref{Approx for p}),(\ref{Approx for conserv}), we
immediately obtain the $C^{0}$-version of (\ref{Travel is 0}) and
(\ref{Correct normal}).

Let us start with the non-tangent case, i.e. with the trajectories
such that $p_{y}(t_{in})$ is bounded away from zero. From
Condition III it follows that the value of
$W_{in}=W_{out}=W(Q=\delta;\epsilon)$ vanishes as $\epsilon
\rightarrow+0$. Hence, by (\ref{Approx for conserv}) the momentum
$p_{y}(t)$ stays bounded away from zero as long as the potential
$W(Q;\epsilon)$ remains small. Choose some small $\nu$, and divide
$N_{\delta}$ into two parts $N_{<}:=\{W:W(Q;\epsilon)\leq\nu\}$
and $N_{>}=\{W:W(Q;\epsilon)>\nu\}$.
First, the trajectory enters $N_{<}$. Since the value of $\frac{d}%
{dt}Q(q)=p_{x}Q_{x}+p_{y}Q_{y}$ is negative and bounded away from
zero in $N_{<}$ (because $Q_{x}$ is small, and $p_{y}$ and $Q_{y}$
are non-zero), the trajectory must reach the inner part $N_{>}$ by
a time proportional to the width of $N_{<}$, which is $O(\delta)$.
Also, we can conclude that if the trajectory leaves $N_{>}$ after
some time $t_{>}$, it must have $p_{y}>0$ and, arguing as above,
we obtain that $t_{out}-t_{in}=O(\delta)+t_{>}$. Let us show that
$t_{>}\rightarrow0$ as $\epsilon\rightarrow+0$. Using
(\ref{Equations of motion y}), the fact that the total variation
of $p_{y}$ is bounded, and Condition IV, we obtain
\[
|t_{>}|\leq\frac{C}{\min_{N_{>}}|W^{\prime}(Q;\epsilon)|}=C\max_{N_{>}%
}|{\mathcal{Q}}^{\prime}(W;\epsilon)|\rightarrow0\;\;\;\mathrm{as}%
\;\;\epsilon\rightarrow+0.
\]
So, in the non-tangent case, the collision time is
$O(\delta+t_{>})$, i.e. it tends to zero indeed.

This result holds true for $p_{y,in}$ bounded away from zero, and
it remains valid for $p_{y,in}$ tending to zero sufficiently
slowly. Hence, we are left with the case where $p_{y,in}$ tends to
zero as $\epsilon\rightarrow0$ (the case of nearly tangent
trajectories). Inside $N_{\delta}$, since $W$ is
monotone by Condition IIc, we have $W(Q;\epsilon)>W_{in}%
=W(\delta;\epsilon)$. Therefore, by (\ref{Approx for conserv}),
$p_{y}(t)$ stays small unless the trajectory leaves $N_{\delta}$
or $t-t_{in}$ becomes larger than a certain bounded away from zero
value. From (\ref{Approx for p}) it follows then that $p_{x}(t)$
remains bounded away from zero. By (\ref{Equations of motion
x}),(\ref{Equations of motion y}),
\[
\dot{Q}:=\frac{d}{dt}Q(q(t);\epsilon)=Q_{x}p_{x}+Q_{y}p_{y}%
\]
so $\dot{Q}$ is small, yet%

\[
\frac{d^{2}}{dt^{2}}Q(q(t);\epsilon)=p_{x}^{T} Q_{xx}p_{x}+2Q_{xy}p_{x}%
p_{y}+Q_{yy}p_{y}^{2}-W^{\prime}(Q;\epsilon)(Q_{x}^{2}+Q_{y}^{2}).
\]
For a non-degenerate tangency, $p_{x}^{T} Q_{xx}p_{x}$ is positive
and bounded away from zero. Therefore, as $p_{y}$ is small and
$W^{\prime}(Q;\epsilon)$ is negative, we obtain that
$\frac{d^{2}}{dt^{2}}Q(q(t);\epsilon)$ is positive and bounded
away from zero for a bounded away from zero interval of time
(starting with $t_{in}$). It follows that
\begin{equation}
Q(q(t);\epsilon)\geq Q(q_{in};\epsilon)+\dot{Q}(t_{in})(t-t_{in}%
)+C(t-t_{in})^{2} \label{qdd}%
\end{equation}
on this interval, for some constant $C>0$. We see from
(\ref{qdd}), that the trajectory has to leave the boundary layer
$N_{\delta}=\{|Q(q;\epsilon
)-Q_{i}|\leq\delta=|Q(q_{in};\epsilon)-Q_{i}|\}$ in a time of
order $O(\dot
{Q}(t_{in}))=O(Q_{x}(q_{in}))+O(p_{y,in})=O(q_{in}-q_{c})+O(p_{y,in})$.
As $q_{in}-q_{c}=O(\sqrt{\delta})$ for a non-degenerate tangency,
we see that the time the nearly-tangent orbit may spend in the
boundary layer is $O(\sqrt{\delta}+p_{y,in})$, i.e. in this case
it tends to zero as well. This completes the proof of Theorem
\ref{Main th0}.

Now we prove Theorem \ref{Main thr} - the $C^{r}$-convergence for
the non-tangent case. Again, divide $N_{\delta}$ into $N_{<}$ and
$N_{>}$ for a small $\nu$ and consider the limit
$\lim_{\delta\rightarrow0}\lim
_{\nu\rightarrow0}\lim_{\epsilon\rightarrow+0}$. As we have shown
above, $\dot{Q}\neq0$ in $N_{<}$, thus we can divide the equations
of motion (\ref{Equations of motion x}), (\ref{Equations of motion
y}) by $\dot{Q}$:
\begin{align}
\frac{dq}{dQ}  &  =\frac{p}{Q_{x}p_{x}+p_{y}Q_{y}}\nonumber\\
\frac{dp}{dQ}  &  =-W^{\prime}(Q;\epsilon)\frac{\nabla Q}{Q_{x}p_{x}%
+p_{y}Q_{y}},\label{Divided equqtion of motion}\\
\frac{dt}{dQ}  &  =\frac{1}{Q_{x}p_{x}+p_{y}Q_{y}}\nonumber
\end{align}
Equations (\ref{Divided equqtion of motion}) can be rewritten in
an integral
form:%
\[
q(Q_{2})-q(Q_{1})=\int_{Q_{1}}^{Q_{2}}F_{q}(q,p)dQ,
\]%
\begin{equation}
p(Q_{2})-p(Q_{1})=-\int_{W(Q_{1})}^{W(Q_{2})}F_{p}(q,p)dW(Q),
\label{Integral form}%
\end{equation}%
\[
t(Q_{2})-t(Q_{1})=\int_{Q_{1}}^{Q_{2}}F_{t}(q,p)dQ,
\]
where $F_{q},F_{p}$ and $F_{t}$ denote some functions of $(q,p)$
which are uniformly bounded along with all derivatives. In
$N_{<}$, the change in $Q$ is bounded by $\delta$ and the change
in $W$ is bounded by $\nu$. Hence, the integrals on the right-hand
side are small. Applying the successive approximation method, we
obtain that the Poincar\'{e} map (the solution to (\ref{Integral
form})) from $Q=Q_{1}$ to $Q=Q_{2}$ limits to the identity map
(along with all derivatives with respect to initial conditions) as
$\delta ,\nu\rightarrow0$. It follows that in order to prove the
theorem, i.e. to prove (\ref{Travel is 0}),(\ref{Correct normal}),
we need to prove
\begin{equation}
\lim_{\nu\rightarrow0}\lim_{\epsilon\rightarrow+0}\bigg\|\bigg(q_{out}%
,t_{out}\bigg)-\bigg(q_{in},t_{in})\bigg)\bigg\|_{C^{r}}=0,
\label{Travel is nu}%
\end{equation}
and
\begin{equation}
\lim_{\nu\rightarrow0}\lim_{\epsilon\rightarrow+0}\bigg\|p_{out}%
-p_{in}+2n(q_{in})\langle
p_{in},n(q_{in})\rangle\bigg\|_{C^{r}}=0,
\label{Correct nu}%
\end{equation}
where $(q_{in},p_{in},t_{in})$ and $(q_{out},p_{out},t_{out})$
correspond now to the intersections of the orbit with the
cross-section $W(Q(q,\epsilon ),\epsilon)=\nu$. By Condition IV,
as $\epsilon\rightarrow0$ the function ${\mathcal{Q}}(W;\epsilon)$
tends to zero uniformly along with all its derivatives in the
region $\nu\leq W\leq H$ for any $\nu$ bounded away from zero.
Therefore, the same holds true for a sufficiently slowly tending
to zero $\nu$ and
$W^{\prime}(Q;\epsilon)=({\mathcal{Q}}^{\prime}(W;\epsilon))^{-1}$
is bounded away from zero in the region $N_{>}$. Hence, by
(\ref{Equations of motion y}), the derivative $\dot{p_{y}}$ is
bounded away from zero as well. Therefore, we can divide the
equations of motion
(\ref{Equations of motion x}),(\ref{Equations of motion x}) by $\frac{dp_{y}%
}{dt}$:%
\begin{equation}
\frac{dq}{dp_{y}}=-\mathcal{Q}^{\prime}(W;\epsilon)\frac{p}{Q_{y}%
},\;\;\;\;\;\frac{dt}{dp_{y}}=-\mathcal{Q}^{\prime}(W;\epsilon)\frac{1}{Q_{y}%
},\;\;\;\;\;\frac{dp_{x}}{dp_{y}}=\frac{Q_{x}}{Q_{y}}, \label{dx-dz equation}%
\end{equation}
where
\begin{equation}
W=H-\frac{1}{2}p^{2}. \label{W from conserv}%
\end{equation}
Condition IV implies that the $C^{r}$-limit as
$\epsilon\rightarrow0$ of
(\ref{dx-dz equation}) is%
\begin{equation}
\frac{d(q,t)}{dp_{y}}=0,\;\;\;\;\;\frac{dp_{x}}{dp_{y}}=\frac{Q_{x}}{Q_{y}}
\label{Cr limit}%
\end{equation}
Since the change in $p_{y}$ is finite and the functions on the
right-hand side of (\ref{dx-dz equation}) are all bounded, the
solution of this system is the $C^{r}$-limit of the solution of
(\ref{dx-dz equation}). From (\ref{Cr limit})
we obtain that in the limit $\epsilon\rightarrow0$ $(q_{in},t_{in}%
)=(q_{out},t_{out})$, so (\ref{Travel is nu}) is proved. Second,
we obtain
from (\ref{Cr limit}) that%
\[
(p_{x,out}-p_{x,in})Q_{y}(q_{in};\epsilon)=(p_{y,out}-p_{y,in})Q_{x}%
(q_{in};\epsilon)
\]
in the limit $\epsilon\rightarrow0$, which, in the coordinate
independent vector notation
\begin{equation}
p_{y}=\langle n(q),p\rangle,\qquad p_{x}=p-p_{y}n(q). \label{eq:normal-py}%
\end{equation}
and by using $(q_{in}%
,t_{in})=(q_{out},t_{out}),$ amounts to the correct reflection
law.
\end{proof}
\end{document}